\documentclass[aip, rsi, reprint, amsmath,amsfonts,amssymb]{revtex4-1}
\usepackage{rotating}
\usepackage{graphicx}

\def\KTH{Nanostructure Physics, Royal Institute of Technology (KTH), Albanova, SE-10791 Stockholm, Sweden}

\begin{document}

\title{Calibrating torsional eigenmodes of micro cantilevers \\
 for dynamic measurement of frictional forces}

    \author{Per-Anders Thor\'en}
    \affiliation{\KTH}

    \author{Riccardo Borgani}
    \affiliation{\KTH}
    
    \author{Daniel Forchheimer}
    \affiliation{\KTH}
    
    \author{David B. Haviland}
    \affiliation{\KTH}

\begin{abstract}
Non-invasive thermal noise calibration of both torsional and flexural eigenmodes is performed on numerous cantilevers of 10 different types.  We show that for all tipless and short-tipped cantilevers,  the ratio of torsional to flexural mode stiffness is given by a constant, times the ratio of their resonant frequencies.  By determining this constant we enable a calibration of the torsional eigenmode, starting from a calibration of the flexural eigenmode.   Our results are well motivated from beam theory and we verify them with finite element simulation.  

\end{abstract}

\maketitle

\section{Introduction}
Calibration is central to the development of quantitative surface analysis with the Atomic Force Microscope (AFM).   Both force and displacement are extracted from the same AFM detector signal, so the calibration is actually two-fold:  the detector output voltage must be converted to tip deflection in nanometers, and this deflection must be related to the force on the tip in newtons.   A method has emerged in the last decade which achieves both of these calibrations from one measurement of the thermal equilibrium Brownian motion of the cantilever near its lowest flexural resonance\cite{Higgins2006,SaderGCI}.  Here we apply this method to both the flexural and the torsional mode.  A simplified method of extracting the torsional calibration from the flexural calibration is found for tipless cantilevers.  We validate our simplified method with finite element modeling. 

Previous work on torsional calibration \cite{Munz2010_calib_review,Palacio2010_calib_review,Alvarez-Asencio2013,Green2004} has focused on determining the stiffness constant $\kappa^\text{stat}$ relating a static torque to static angular twist about the major axis of the cantilever.  This static stiffness constant is sufficient if one is only interested in measuring lateral tip-surface force in static equilibrium.  However, the currently popular calibration methods are actually rooted in a determination of the dynamic stiffness of a particular vibration eigenmode of the cantilever.  Dynamic and static stiffness are not the same, and to determine the latter from the former, one must perform finite element simulation of the particular beam geometry in question. In this work we focus on calibration for dynamic measurement of frictional force \cite{Thoren2016} where static stiffness is irrelevant and the mode stiffness is one of three necessary calibration constants.

The calibration constants specify a transduction of force on the AFM tip, originating at the 10 nanometer scale, to a deformation of a cantilever at the 10 micrometer scale.  The transition from nanometer to micrometer scale relies on a model for the continuum mechanics of the cantilever in terms of its normal modes of vibration.  Each mode corresponds to a resonance modeled by a damped harmonic oscillator (DHO) with a linear response function $\hat{\chi}(\omega)$ that relates the frequency components of the tip deflection $\hat{d}(\omega)$ to those of the force $\hat{F}_\text{TS}(\omega)$ acting on the tip,
\begin{equation}
\hat{d}(\omega) = \hat{\chi}(\omega) \hat{F}_\text{TS}(\omega) .
\end{equation}
The DHO linear response function
\begin{equation}
\hat{\chi}(\omega) = \frac{1}{k} \left[ 1 + i \frac{\omega}{\omega_0 Q} -  \frac{\omega^2}{\omega_0^2} \right]^{-1}
\label{eq:chi}
\end{equation}
is specified with three constants: the mode stiffness $k$, resonant frequency $\omega_0$ and quality factor $Q$; or equivalently the stiffness, effective mass $m=k/\omega_0^2$ and the damping coefficient $\eta = \sqrt{km}/Q$, the latter being the dissipative force constant of our linear model. Thus, calibration involves the determination of these three constants of the DHO model. 

This reduction to a simple DHO model loses accuracy when the response $\hat{d}(\omega)$ is spread over a frequency band much wider than one resonance, or when significant components of $\hat{d}(\omega)$ occur near another resonance.  In these cases a more accurate response function could be modeled with a superposition of additional eigenmodes, each with its own $\hat{\chi}(\omega)$ and three calibration constants. Thus, quantitative AFM is greatly simplified when weak tip-surface forces perturb high-$Q$ resonance, such that all motion is well-confined to only one mode.  

Sader {\em et al.} proposed a method for calibrating the stiffness of the fundamental (lowest frequency) flexural eigenmode using knowledge of its hydrodynamic damping in a fluid of arbitrary density and viscosity \cite{Sader1998,Sader2005_scaling}. The method was generalized to thin cantilevers of arbitrary plane view, where a simplifying approximation to the hydrodynamic function required only one 'Sader constant', unique to that geometry \cite{Sader2012}.  Recently a web-based global calibration initiative \cite{SaderGCI} was launched with the aim of using big-data to more accurately determine the hydrodynamic function (Sader constant) for numerous cantilevers. The Sader method gives the dynamic mode stiffness $k$, using the quality factor $Q$ and resonant frequency $\omega_0$ as inputs.  

$Q$ and $\omega_0$ are determined by measuring the total noise power at the output of the detector and fitting a theoretical expression consisting of two independent noise sources,
\begin{equation}
S_\text{tot} = S_\text{det} + \alpha^2 S_\text{cant}(\omega) \hspace{4pt}\left[\mathrm{\frac{V^2}{Hz}}\right]
\label{Eq:Stot}
\end{equation}
where  $S_\text{det}$ is a frequency-independent background detector noise, $\alpha$ [V/nm] the calibration constant converting nanometers of tip deflection to detector voltage, and $S_\text{cant}(\omega)$ the thermal fluctuations of the cantilever.  The latter is given by the fluctuation-dissipation theorem applied to the linear response function of the DHO model eq.~\ref{eq:chi}. 
\begin{equation}
S_\text{cant}(\omega) = \frac{2 k_\text{B}T}{\omega}\mathrm{Im}\left[  \hat{\chi}(\omega) \right] = 2 k_\text{B}T \eta \vert \hat{\chi} \vert^2 \hspace{4pt}\left[\mathrm{\frac{nm^2}{Hz}}\right] .
\label{Eq:FDT}
\end{equation}
  
If one also fits the magnitude of $S_\text{cant}$, the measured temperature together with the value of $k$ given by the Sader method allow for a determination of $\alpha$.\cite{Higgins2006}  Thus, a full calibration is achieved in a frequency band near resonance, from one simple noise measurement.  This is highly advantageous as it does not destroy the sharp tip by pushing on a hard surface, as needed to calibrate $\alpha$ against the AFM scanner.  The calibration of both force and deflection are then traceable to one experimental procedure which is easily performed on any cantilever and completely independent of the scanner calibration.

\section{Torsional calibration}

\begin{figure}
\includegraphics{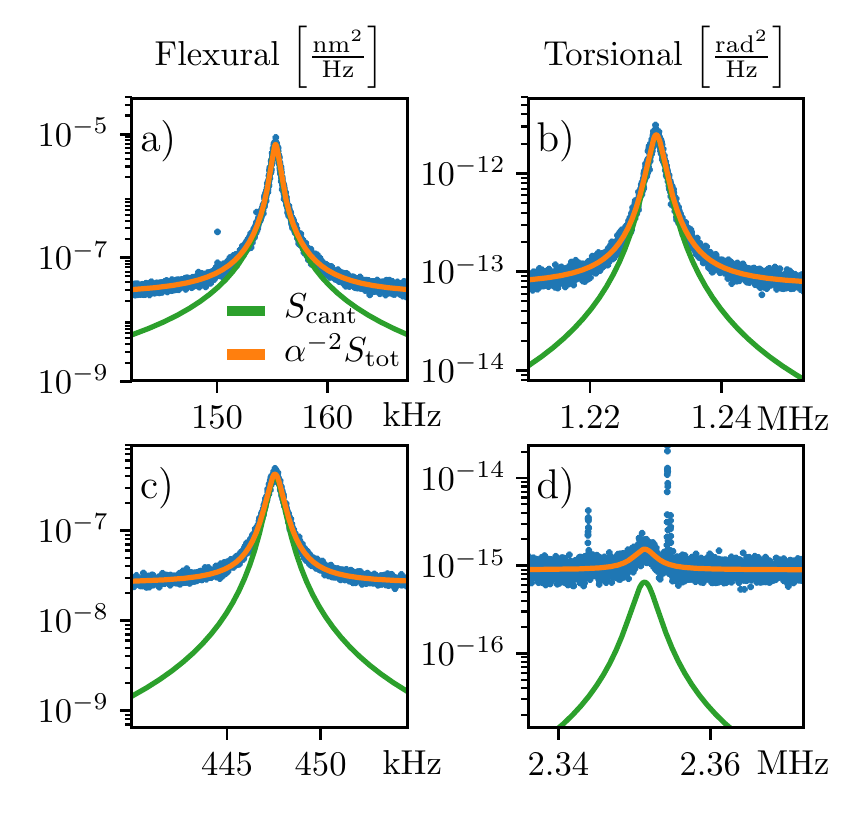}
\caption{ \textbf{Flexural and torsional noise.}   
Measured thermal Brownian motion power spectra for the lowest flexural (a and c) and torsional (b and d) eigenmodes of a soft cantilever (a and b) and stiff cantilever (c and d).  The orange curves are found by fitting the theory Eq.~\ref{Eq:Stot} to the measured noise. The green curves show the cantilever contribution to the total noise. In (d) more averaging is required to resolve the torsional cantilever noise, which drops below the detector noise for the stiffer cantilever. Spurious signals are also observed on either side of resonance.
\label{fig:noise}
}
\end{figure}

The Sader method has been widely applied to the lowest flexural eigenmode but less so to the torsional mode.   The theory of the method was generalized to torsional eigenmodes, \cite{Green2004,anthony_sader_bookchapter} but its application is complicated by difficulties in measuring the thermal fluctuations near torsional resonance, especially for stiff beams\cite{Green2002}.  On resonance the flexural motion noise $\alpha^2 S_\text{cant}$ can be a factor of 100 or more greater than the detector noise $S_\text{det}$ (see fig.~\ref{fig:noise}a).  For the same cantilever and detector, torsional noise is significantly lower in relation to detector noise (see fig.~\ref{fig:noise}b).  With stiffer levers the torsional noise often falls below the detector noise, making accurate measurement of the torsional quality factor rather difficult (see fig.~\ref{fig:noise}d).

Here we propose an alternative to the torsional Sader method which does not rely on measurement of torsional fluctuations.  The idea is to bootstrap from a non-invasive calibration of the lowest flexural mode, and extract the torsional mode stiffness using only the torsional resonant frequency as input.  We are motivated by beam theory~\cite{Lifshitz-Landau,anthony_sader_bookchapter} which shows that all material constants fall away when considering the ratio of flexural and torsional eigenfrequencies.  Assuming only a uniform long, thin beam, we arrive at an expression for the torsional stiffness $\kappa_\text{t}$ [N m/rad] in terms of the flexural stiffness $k_\text{f}$ [N/m] which depends only on the ratio of resonant frequencies and the width of the beam $b$ (see Appendix).
\begin{equation}
\kappa_\text{t} = \frac{k_\text{f} b^2}{6}  \left(\frac{\omega_\text{0t}^\text{(vac)}}{\omega_\text{0f}^\text{(vac)}}\right)^2.
\label{eq:k_t}
\end{equation}
Here the eigenfrequencies are vacuum values, without the added mass-loading of a surrounding fluid, and the subscripts t and f refer to torsional and flexural respectively. Taking into account the hydrodynamic load and damping when the beam is moving in a viscous fluid, theory gives (see Appendix and Refs.~\cite{Green2002,Green2004}),
\begin{equation}
\kappa_\text{t} =  C k_\text{f} b^2\left(\frac{\omega_\text{0t}}{\omega_\text{0f}}\right)^2 .
\label{eq:k_t_corr}
\end{equation}
The correction factor
\begin{equation}
C = \left(  \frac{Q_\text{t}}{Q_\text{f}} \right)  \frac{\Gamma_\text{t}(\mathrm{Re}_\text{t})}{\Gamma_\text{f}(\mathrm{Re}_\text{f})}.,
\label{eq:corr}
\end{equation}
depends on ratios of quality factors $Q_\text{t(f)}$ and hydrodynamic functions $\Gamma_\text{t(f)}(\mathrm{Re}_\mathrm{t(f)})$ of the torsional and flexural modes.  The latter are functions of the Reynolds numbers\cite{Sader1998,Green2002},
\begin{equation}
\mathrm{Re}_\text{t(f)} = \frac{\rho \omega_\text{0t(f)} b^2}{4 \mu}
\label{Re}
\end{equation}
where $\rho$ and $\mu$ are the density and viscosity of the ambient fluid.

Below we present experiments in room temperature air.  Our analysis shows that for tipless and short-tipped cantilevers, the correction factor $C$ is independent of the ratio of Reynolds numbers,
\begin{equation}
\frac{\mathrm{Re}_\text{t}}{\mathrm{Re}_\text{f}} = \frac{\omega_\text{0t}}{\omega_\text{0f}} \equiv \widetilde{\omega}
\end{equation}
 and  only slightly below its ideal vacuum value $C=\frac{1}{6}$.  This observation indicates that either with or without fluid, the essential dimensionless parameter for determining the ratio of beam stiffnesses is simply $\widetilde{\omega}$.  However, for cantilevers with longer tips we find that $C$ is further suppressed below its ideal vacuum value.  Nevertheless, knowing $C$ for a particular AFM probe allows for determining $\kappa_\text{t}$ from a flexural calibration and a measurement of the torsional resonant frequency.  The latter can be easily measured by a driven frequency sweep, without measuring torsional thermal fluctuations.  However, the method would still require an independent measurement of the detector constant $\alpha_t$ for complete calibration.

\section{Results \& Discussion}

Noise measurements are made on 10 standard cantilevers of different types, as specified in table~\ref{tab:cant}.  We measure many cantilevers of each type using two different AFMs having optical beam-deflection detectors (JPK NanoWizard 3 and Bruker Dimension Icon).  Figure~\ref{fig:noise} shows examples of noise measurements on two different cantilevers, and the result of fitting eqs.~\eqref{Eq:Stot} and \eqref{Eq:FDT} to the data.  From this fitting for each cantilever we extract $\omega_0$ and $Q$ for the lowest flexural and torsional eigenmode.  A worst-case fit is shown in fig.~\ref{fig:noise}d for the stiffest cantilever No.9, where one can clearly see that the fit to determine $Q_\text{t}$ is less reliable than that for $Q_\text{f}$. 

\begin{table}
    \begin{tabular}{| c | c | c | c | c | c | c | c |}
    \hline
    No. & Name      & $h_\text{tip}$  & $L$        & $b$       & $\sim f_\text{0f}$   & $\sim k_\text{f}$  & Plane view     \\ 
        &           &[$\mu$m]         & [$\mu$m]   &[$\mu$m]   & [kHz]                & [Nm$^{-1}$]        &                \\ \hline
    1   & NO-CAL    & 0               & 397        & 29        & 18                   & 0.16               & Rectangular    \\ \hline
    2   & NO-CAL    & 0               & 197        & 29        & 71                   & 1.3                & Rectangular    \\ \hline
    3   & NO-CAL    & 0               & 97         & 29        & 293                  & 10.4               & Rectangular    \\ \hline \hline
    4   & ORC8-A    & 2-3.5           & 100        & 40        & 71                   & 0.73               & Rectangular    \\ \hline
    5   & ORC8-C    & 2-3.5           & 100        & 20        & 68                   & 0.38               & Rectangular    \\ \hline
    6   & ORC8-D    & 2-3.5           & 200        & 20        & 18                   & 0.05               & Rectangular    \\ \hline \hline
    7   & Tap150    & 10-15           & 125        & 30        & 150                  & 5                  & Picket fence   \\ \hline
    8   & Tap300    & 10-15           & 125        & 35        & 300                  & 40                 & Picket fence   \\ \hline
    9   & Tap525    & 10-15           & 125        & 40        & 525                  & 200                & Picket fence   \\ \hline \hline
    10  & MPP33120  & 10-20           & 450        & 40        & 40                   & 5                  & Picket fence   \\ \hline
    \end{tabular}
    \caption{Overview of cantilevers used in this study. The tip height $h_\text{tip}$, length $L$ and width $b$ are those given by the manufacturer. The flexural resonance frequency and stiffness are nominal values. Actual measured values on many probes of each type are used in fig.~\ref{fig:measurements}, where the cantilever type is distinguished using colored symbols.
    \label{tab:cant}}
\end{table}

\begin{figure}
\includegraphics{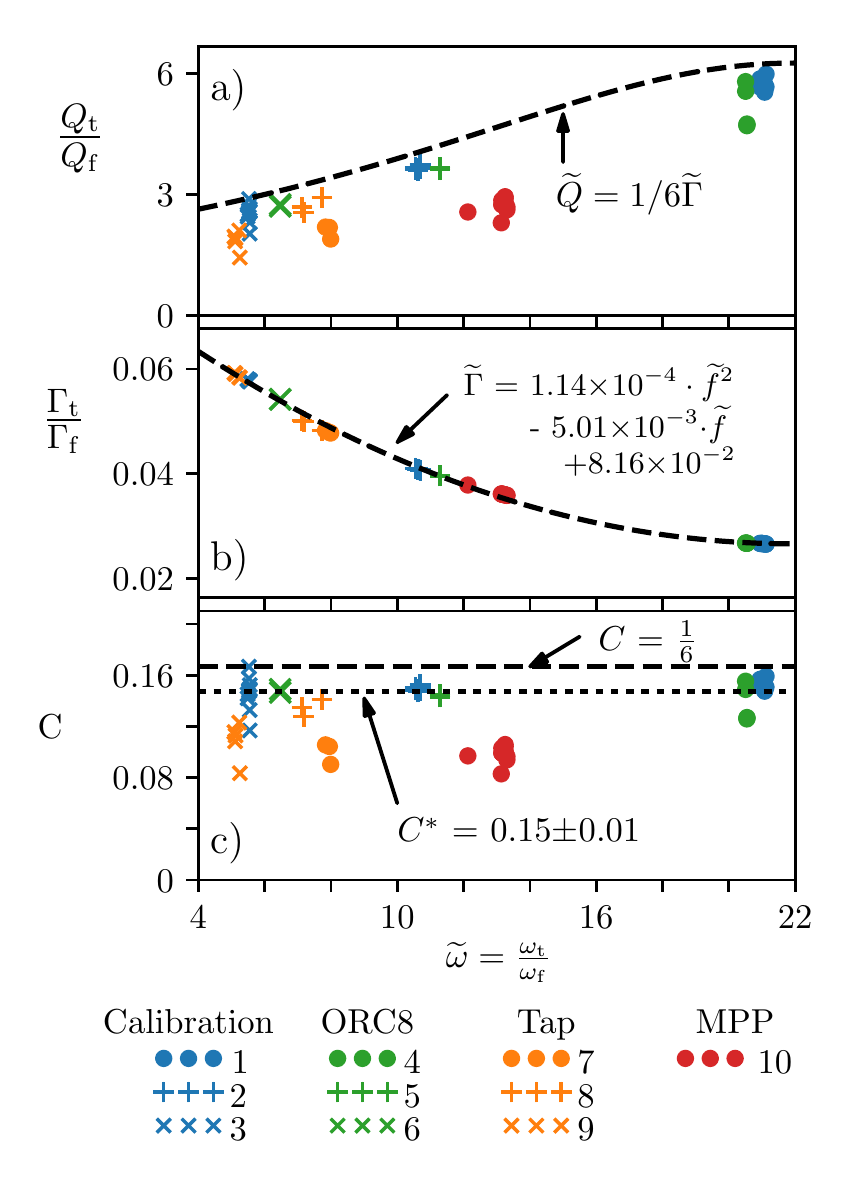}
\caption{ \textbf{Calculation of the C-factor.} 
(a) Measurement of the ratio of torsional to flexural quality factors ($\widetilde{Q} = Q_t/Q_f$) for different cantilevers.  Dashed line is $\widetilde{Q} = \frac{1}{6\widetilde{\Gamma}}$, with $\widetilde{\Gamma}$ from panel fig.~\ref{fig:measurements}~b. (b) Calculation of the ratio of torsional to flexural hydrodynamic function ($\widetilde{\Gamma} = \Gamma_t/\Gamma_f$), with a parabolic fit.  (c) Calculated C-factor for different cantilevers.  The $C$-factor predicted by theory, $C=\frac{1}{6}$, is marked with dashes and a fit to cantilevers No. 1-6 is marked with dots.  Common x-axis is torsional to flexural resonance frequency ratio, $\widetilde{\omega} = \omega_\mathrm{t}/\omega_\mathrm{f}$.}
\label{fig:measurements}
\end{figure}

In fig.~\ref{fig:measurements} we plot for each beam, the ratio $Q_\text{t}/Q_\text{f}$ vs. $\widetilde{\omega}$.  With the fitted value of $\omega_\text{0t(f)}$, $b$ given in table~\ref{tab:cant}, the density $\rho=1.18$~kg/m$^3$ and dynamic viscosity $\mu=1.86\cdot10^{-5}$~kg/(ms) of air, we calculate the flexural and torsional Reynolds number for each beam.  With the Reynolds numbers we use the theoretical expressions in Sader \textit{et al.}\cite{Sader1998} and Green \textit{et al.}\cite{Green2002} to calculate the ratio of the torsional and flexural hydrodynamic functions, plotted in fig.~\ref{fig:measurements}b. It is interesting to note that this quantity falls on a smooth curve when plotted vs. $\widetilde{\omega}$, for cantilevers spanning a factor of two in $b$ and a factor of 30 $\omega_\text{0f}$.  We fit this smooth curve to a polynomial of degree 2 to find the coefficients given in fig.~\ref{fig:measurements}b.

Using the ratios of quality factors and hydrodynamic functions, we use eq.~\ref{eq:corr} to plot the correction factor $C$, shown in fig.~\ref{fig:measurements}c.  The dashed line shows the ideal vacuum value $C=\frac{1}{6}$.  The three tipless cantilevers (No.~1-3, red data points), and the ORC cantilevers with short tips (No.~4-6, blue data points), all fall slightly below this vacuum value.  The scatter in the data at small $\widetilde{\omega}$ comes from stiff beams where $Q_\text{t}$ is difficult to determine, as discussed above.   At larger $\widetilde{\omega}$ the scatter is considerably less, and over the full range of $\widetilde{\omega}$ studied a least square fit of C for cantilevers No.~1-6 gives $C=0.15 \pm 0.01$. Another source of error is the uncertainty in the width $b$.

For the beams with larger tips, we see that $C$ is further suppressed.  This observation can be explained by fluid flow being affected by the tip, a pyramid of height $h_\text{tip}$ normal to the beam width.  For flexural motion the tip is in the 'shadow' of the beam and flow is minimally perturbed.  Such is not the case for torsional motion where the protruding tip can significantly affect the flow.  Our analysis indicates that the torsional Sader method can not be trusted for beams with tall tips.  Nevertheless, if the location and shape of the tip is reproducible in the cantilever manufacturing process, one could imagine a torsional Sader method with an experimentally determined value of $C$.  

We note that our analysis to arrive at $C$ does not rely on any knowledge of the detector constants $\alpha_\text{t}$ and $\alpha_\text{f}$.  These can be determined from the magnitude of the thermal fluctuation, or geometric methods with controlled tilting of the AFM head\cite{Pettersson2007}.

\section{Finite element modeling}

Equation \ref{eq:k_t} applies to a tipless cantilever with length $L$ much greater than width $b$, which is not the case for all of our beams.  To understand how our results might be influenced by the plane-view aspect ratio $L/b$, we perform finite element modeling (FEM) on cantilevers No.~1-3 using COMSOL. We model beams with $L$ and $b$ given in Table \ref{tab:cant} using tabulated values for the bulk modulus $E=170$~GPa, density $\rho_c=2329$~kg/m$^3$ and Poisson ratio $\nu=0.28$. We begin with an eigenmode analysis to find the 6 lowest eigenmodes, including flexural, torsional and lateral modes, where the latter is side-ways bending of the beam in the plane defined by $L$ and $b$. The beam thickness is then adjusted such that the lowest flexural eigenfrequency is very close to the nominal value given in Table \ref{tab:cant}.

We determine the static stiffnesses by applying a known static force (torque) to the free end, to find the slope of the deflection vs. force (torque) curve.   The dynamic stiffness is found by applying a known periodic drive plus a weak viscous damping force (torque) to the free end, while sweeping the drive frequency through the eigenfrequency of interest.  We then fit the simulated high Q response curve to eq.~\eqref{eq:chi} to determine $k$ ($\kappa$).  Having thus determined the mode stiffnesses and resonant frequencies, we calculate the correction factor $C$.  Table \ref{tab:FEM} tabulates the relevant results.

\begin{figure}
\includegraphics[width=\columnwidth]{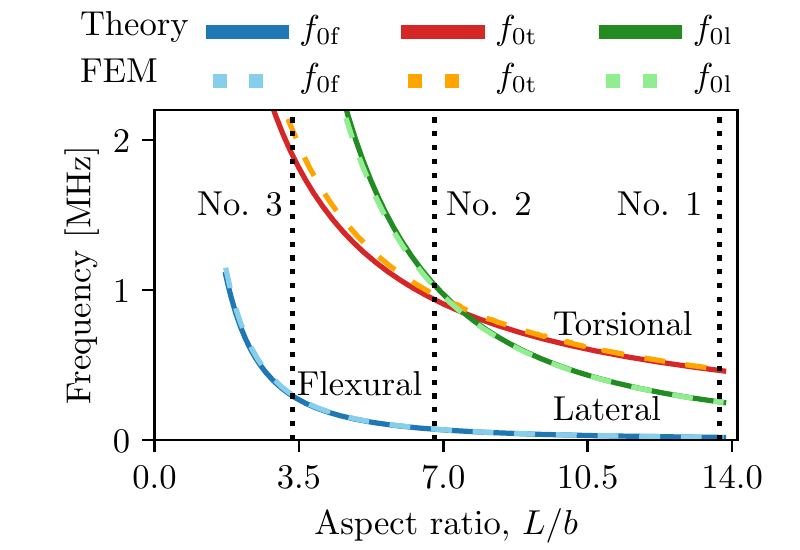}
\caption{ \textbf{Resonance frequencies vs cantilever length.} The first flexural $f_\mathrm{0f}$, torsional $f_\mathrm{0t}$ and lateral $f_\mathrm{0l}$ resonance frequencies as a function of cantilever length.  The solid lines are from beam theory, \textit{i.e.} eqs.~\eqref{eq:w0vac_beam} and \eqref{eq:w0t_vac_beam} in the Appendix; the dashed lines are obtained with FEM simulations.  The three dotted vertical lines mark the aspect ratio of cantilevers No. 1--3.   The width and thickness are kept constant, at $b=29~\mu$m and $t=2~\mu$m.} 
\label{fig:crossings}
\end{figure}

The correction factor of the shortest beam is quite close to the ideal beam theory $C=0.166...$, whereas the longest beam deviates substantially and the intermediate beam is quite far off.  This is a curious result as we would expect that beam theory is approached in the limit $L \gg b$.  An explanation for this discrepancy is found by plotting the lowest eigenmode frequency vs. $L/b$ for the flexural, torsional and lateral modes, as shown in fig.~\ref{fig:crossings}.   For each mode we plot the beam theory expressions (eqs.\eqref{eq:w0vac_beam} and \eqref{eq:w0t_vac_beam} in the Appendix), as well as the result from FEM simulations.  

Beam theory agrees very well with FEM simulation over the range of $L/b$ studied, with slight deviation in the torsion mode at lower $L/b$. For the shortest cantilever No.~3 we see that the lateral mode is very stiff, with  much higher frequency than the torsional or flexural.  For the longest cantilever No.~1 the lateral mode becomes softer than the torsional mode, and the intermediate cantilever No.~2 is very close to a crossing of the torsional and lateral modes.  When the modes become close in frequency, any small deviation from ideal geometry, for example the addition of a tip, will result in mode coupling.  

Indeed, we frequently observe two modes when calibrating non-ideal beams of intermediate $L/b$ and it can be difficult to determine which of the two is the torsional mode.  Torsional AFM is difficult with the longest beams because tip-surface forces can easily excite the lateral mode.  Long beams also have low flexural stiffness, making it difficult to regulate the load force without jump-to-contact instabilities.  Thus, stiff, short beams are preferred for dynamic friction measurements, in spite of difficulties in measuring torsional fluctuations.

\begin{table}
    \begin{tabular}{| c || c | c | c || c | c | c || c |}
    \hline
    No. & $f_\text{0f}$      & stat. $k_\text{f}  $  & dyn. $k_\text{f}$    & $f_\text{0t}$    & stat. $\kappa_t$        & dyn. $\kappa_t$          & $C$\\ 
        & [kHz]             & [Nm$^{-1}$]  & [Nm$^{-1}$]  & [kHz]    & [Nm]               & [Nm]                 &                      \\ \hline
    1   & 17.5       & 0.16         & 0.17         & 471      & 1.31$\cdot10^{-8}$ & 1.76$\cdot10^{-8}$   & 0.1752                 \\ \hline
    2   & 71.6       & 1.31         & 2.07         & 969      & 2.67$\cdot10^{-8}$ & 3.82$\cdot10^{-8}$   & 0.1194                 \\ \hline
    3   & 299        & 11.2         & 14.5         & 2062     & 5.67$\cdot10^{-8}$ & 9.75$\cdot10^{-8}$   & 0.1683                 \\ \hline
    \end{tabular}
    \caption{FEM simulations of cantilevers No. 1-3.  The $C$-factor is calculated from the dynamical $k_\mathrm{f}$ and $\kappa_\mathrm{t}$.}
    \label{tab:FEM}
\end{table}

\section{Conclusions}

We measured thermal fluctuations near flexural and torsional resonances on a wide variety of cantilevers in room temperature air.  We applied the hydrodynamic method of Sader {\it et al.} and Green {\it et al.} to extract the flexural and torsional mode stiffnesses. Taking the ratio of torsional to flexural stiffness we found that eq.~\eqref{eq:k_t_corr} holds for tipless and short-tipped beams, with a least-square fit value $C=0.15\pm0.01$.  Using eq.\eqref{eq:k_t_corr} to calculate $\kappa_\text{t}$  we avoid the uncertainty of an inaccurate model of the torsional damping.  One interpretation of this approach is that the flexural calibration is used to determine the beam properties (material parameters and thickness), and beam theory gives torsional stiffness.  Equation~\eqref{eq:k_t_corr} is therefore a more powerful approach than many competing methods in the sense that it does not rely on either exact material values or an accurate damping model.  For shorter, stiffer beams where torsional noise is difficult to measure, torsional stiffness can be found from noise calibration of the flexural mode, together with the torsional resonance frequency found from a driven frequency sweep.  We used FEM to show that eq.\eqref{eq:k_t_corr} holds for these short, stiff beams.  Such beams are well-suited for dynamic friction measurements because the torsional mode is well separated from the lateral mode.

\section*{Acknowledgements}
We gratefully acknowledge financial support from the Swedish Research Council (VR) and the Knut and Alice Wallenberg Foundation.  We also acknowledge helpful discussions with John E. Sader and C. Anthony van Eysden.

\section*{Bibliography}

\begin{thebibliography}{10}

\bibitem{Higgins2006}
M.~J. Higgins, R.~Proksch, John~Elie Sader, M.~Polcik, S.~{Mc Endoo}, J.~P.
  Cleveland, and S.~P. Jarvis.
\newblock {Noninvasive determination of optical lever sensitivity in atomic
  force microscopy}.
\newblock {\em Review of Scientific Instruments}, 77(1):013701, 2006.

\bibitem{SaderGCI}
John~Elie Sader, Riccardo Borgani, Christopher~T. Gibson, David~B. Haviland,
  Michael~J. Higgins, Jason~I. Kilpatrick, Jianing Lu, Paul Mulvaney,
  Cameron~J. Shearer, Ashley~D. Slattery, Per-Anders Thor{\'{e}}n, Jim Tran,
  Heyou Zhang, Hongrui Zhang, and Tian Zheng.
\newblock {A virtual instrument to standardise the calibration of atomic force
  microscope cantilevers}.
\newblock {\em Review of Scientific Instruments}, 87(9), 2016.

\bibitem{Munz2010_calib_review}
Martin Munz.
\newblock {Force calibration in lateral force microscopy: a review of the
  experimental methods}.
\newblock {\em Journal of Physics D: Applied Physics}, 43(6):063001, 2010.

\bibitem{Palacio2010_calib_review}
Manuel L.~B. Palacio and Bharat Bhushan.
\newblock {Normal and Lateral Force Calibration Techniques for AFM
  Cantilevers}.
\newblock {\em Critical Reviews in Solid State and Materials Sciences},
  35(2):73--104, 2010.

\bibitem{Alvarez-Asencio2013}
R~Álvarez-Asencio, E~Thormann, and Mark~W. Rutland.
\newblock {Note: Determination of torsional spring constant of atomic force
  microscopy cantilevers: Combining normal spring constant and classical beam
  theory.}
\newblock {\em The Review of scientific instruments}, 84(9):096102, September
  2013.

\bibitem{Green2004}
Christopher~P. Green, Hadi Lioe, Jason~P. Cleveland, Roger Proksch, Paul
  Mulvaney, and John~Elie Sader.
\newblock {Normal and torsional spring constants of atomic force microscope
  cantilevers}.
\newblock {\em Review of Scientific Instruments}, 75(6):1988, 2004.

\bibitem{Thoren2016}
Per-Anders Thor\'{e}n, Astrid~S. de~Wijn, Riccardo Borgani, Daniel Forchheimer,
  and David~B. Haviland.
\newblock {Imaging high-speed friction at the nanometer scale}.
\newblock {\em Nature Communications}, 7:13836, 2016.

\bibitem{Sader1998}
John~Elie Sader.
\newblock {Frequency response of cantilever beams immersed in viscous fluids
  with applications to the atomic force microscope}.
\newblock 84(1):64--76, 1998.

\bibitem{Sader2005_scaling}
John~E. Sader, Jessica Pacifico, Christopher~P. Green, and Paul Mulvaney.
\newblock {General scaling law for stiffness measurement of small bodies with
  applications to the atomic force microscope}.
\newblock {\em Journal of Applied Physics}, 97(12), 2005.

\bibitem{Sader2012}
John~Elie Sader, Julian~a Sanelli, Brian~D Adamson, Jason~P Monty, Xingzhan
  Wei, Simon~a Crawford, James~R Friend, Ivan Marusic, Paul Mulvaney, and
  Evan~J Bieske.
\newblock {Spring constant calibration of atomic force microscope cantilevers
  of arbitrary shape.}
\newblock {\em The Review of scientific instruments}, 83(10):103705, October
  2012.

\bibitem{anthony_sader_bookchapter}
Cornelis~Anthony van Eysden and John~Elie Sader.
\newblock {\em Frequency Response of Cantilever Beams Immersed in Viscous
  Fluids}, chapter~2, pages 29--53.
\newblock Wiley-Blackwell, 2015.

\bibitem{Green2002}
Christopher~P. Green and John~Elie Sader.
\newblock {Torsional frequency response of cantilever beams immersed in viscous
  fluids with applications to the atomic force microscope}.
\newblock {\em Journal of Applied Physics}, 92(10):6262, 2002.

\bibitem{Lifshitz-Landau}
L.~D. Landau and E.~M. Lifshitz.
\newblock {\em Theory of Elasticity}.
\newblock Pergamon, 1959.

\bibitem{Pettersson2007}
Torbj{\"{o}}rn Pettersson, Niklas Nordgren, Mark~W. Rutland, and Adam Feiler.
\newblock {Comparison of different methods to calibrate torsional spring
  constant and photodetector for atomic force microscopy friction measurements
  in air and liquid}.
\newblock {\em Review of Scientific Instruments}, 78(9), 2007.

\bibitem{VanEysden2006}
Cornelis~A. {Van Eysden} and John~Elie Sader.
\newblock {Resonant frequencies of a rectangular cantilever beam immersed in a
  fluid}.
\newblock {\em Journal of Applied Physics}, 100(11):114916, 2006.

\bibitem{Sader1999}
John~Elie Sader, James W.~M. Chon, and Paul Mulvaney.
\newblock {Calibration of rectangular atomic force microscope cantilevers}.
\newblock {\em Review of Scientific Instruments}, 70(10):3967, 1999.

\end{thebibliography}

\clearpage
\appendix*
\section*{Appendix 1}
\setcounter{equation}{0}


The resonance frequency of the lowest flexural eigenmode in vacuum is given by eq.~(12) in Sader \textit{et al.} 1998\cite{Sader1998}:
\begin{equation}
    \omega_{0f}^\text{(vac)} = \frac{C_1^2}{L^2}\sqrt{\frac{EI}{\rho_c b h}}
    \label{eq:w0vac_beam}
\end{equation}
where $C_1~\approx~1.875104...$ and $I = bh^3/12$ is the second moment of area.
The flexural mode stiffness is:
\begin{equation}
    k_\mathrm{f} \equiv \frac{C_1^4EI}{4L^3}
    \label{eq:k_beam}
\end{equation}
giving
\begin{equation}
    \omega_{0f}^\text{(vac)} = 2\sqrt{\frac{k_f}{\rho_c L b h}}.
    \label{eq:w0vac_beam2}
\end{equation}
Similarly for the lowest torsional eigenmode,  eq.~(15) in van Eysden \textit{et al.} 2006\cite{VanEysden2006} is:
\begin{equation}
    \omega_{0t}^\text{(vac)} = \frac{\pi}{2L}\sqrt{\frac{GK}{\rho_c I_p}}
    \label{eq:w0t_vac_beam}
\end{equation}
where $K=bh^3/3$ and $I_\mathrm{p}=b^3h/12$ for a thin rectangular beam. The torsional mode stiffness is
\begin{equation}
    \kappa_t = \frac{\pi^2GK}{8L}.
    \label{eq:kt_beam}
\end{equation}
giving
\begin{equation}
    \omega_{0t}^\text{(vac)} = \frac{2\sqrt{6}}{b}\sqrt{\frac{\kappa_t}{\rho_c L b h}}.
    \label{eq:w0t_vac_beam2}
\end{equation}
Taking the ratio of eqs.~\eqref{eq:w0vac_beam2} and \eqref{eq:w0t_vac_beam2} gives eq.~\eqref{eq:k_t} in the main text:
\begin{equation}
\kappa_\mathrm{t} = \frac{k_\mathrm{f}b^2}{6}\left(\frac{\omega_\mathrm{0t}^\text{(vac)}}{\omega_\mathrm{0f}^\text{(vac)}}\right)^2.
\label{eq:k_ratio_vac}
\end{equation}

The vacuum frequencies, $\omega_{0f}^\text{(vac)}$ and $\omega_{0t}^\text{(vac)}$, are related to those measured in a viscous fluid, $\omega_{0f}$ and $\omega_{0t}$, via eqs.~(2) and (3) in Sader \textit{et al.} 1999\cite{Sader1999}:
\begin{equation}
    \omega_\mathrm{0f}^\text{(vac)} = \omega_\mathrm{0f} \left[ 1 + \frac{\pi \rho b}{4 \rho_\mathrm{c} h}\Gamma_\mathrm{r}(\mathrm{Re}_\mathrm{f})\right]^{\frac{1}{2}}
    \label{eq:w0vac_sader}
\end{equation}
\begin{equation}
    \rho_c h = \frac{\pi \rho b}{4}\left[ Q_\mathrm{f} \Gamma_i^f(\mathrm{Re}_\mathrm{f}) - \Gamma_r^f(\mathrm{Re}_\mathrm{f})\right]
    \vspace{4pt}
    \label{eq:w0vac_sader2}
\end{equation}
and eqs.~(15) and (16) from Green \textit{et al.} 2004\cite{Green2004}:
\begin{equation}
    \omega_\text{0t}^\text{(vac)} = \omega_t \left[ 1 + \frac{3\pi \rho b}{2 \rho_c h}\Gamma_r^t(Re_t) \right]^{\frac{1}{2}}
    \label{eq:w0t_vac_green}
\end{equation}
\begin{equation}
    \rho_c h = \frac{3}{2} \pi \rho b \left[Q_t \Gamma_i^t(Re_t) - \Gamma_r^t(Re_t)\right].
    \label{eq:w0t_vac_green2}
\end{equation}
Combining eqs.~\eqref{eq:w0vac_sader} -- \eqref{eq:w0t_vac_green2} and inserting into eq.~\eqref{eq:k_ratio_vac} gives eqs.~\eqref{eq:k_t_corr} and \eqref{eq:corr} of the main text:
\begin{equation}
\kappa_\mathrm{t} = k_\mathrm{f}b^2\left(\frac{\omega_\mathrm{0t}}{\omega_\mathrm{0f}}\right)^2 \underbrace{\frac{Q_\mathrm{t}}{Q_\mathrm{f}}\frac{\Gamma_\mathrm{t}(\mathrm{Re}_\mathrm{t})}{\Gamma_\mathrm{f}(\mathrm{Re}_\mathrm{f})}}_{\equiv C}.
\label{eq:k_ratio_novac}
\end{equation}

\end{document}